\PassOptionsToPackage{table,xcdraw}{xcolor}

\documentclass[sigconf]{acmart}

\AtBeginDocument{%
  \providecommand\BibTeX{{%
    \normalfont B\kern-0.5em{\scshape i\kern-0.25em b}\kern-0.8em\TeX}}}

\usepackage{makecell}
\usepackage{cleveref}
\usepackage{amsthm}
\usepackage{bm}
\usepackage[table,xcdraw]{xcolor}
\usepackage{colortbl}
\usepackage{amsmath}
\usepackage{multirow}
\usepackage{hyperref}
\usepackage{graphicx}
\usepackage{subfigure}
\usepackage{enumitem}
\usepackage{booktabs}
\usepackage{bbding}
\usepackage{color}
\usepackage{xcolor}
\usepackage{booktabs}

\usepackage{listings}
\usepackage{natbib}
\usepackage{xspace} 
\usepackage[ruled,vlined,linesnumbered]{algorithm2e}
\usepackage[table]{xcolor}

\usepackage{etoolbox}
\makeatletter
\AfterEndEnvironment{algorithm}{\let\@algcomment\relax}
\AtEndEnvironment{algorithm}{\kern2pt\hrule\relax\vskip3pt\@algcomment}
\let\@algcomment\relax
\newcommand\algcomment[1]{\def\@algcomment{\footnotesize#1}}
\renewcommand\fs@ruled{\def\@fs@cfont{\bfseries}\let\@fs@capt\floatc@ruled
  \def\@fs@pre{\hrule height.8pt depth0pt \kern2pt}%
  \def\@fs@post{}%
  \def\@fs@mid{\kern2pt\hrule\kern2pt}%
  \let\@fs@iftopcapt\iftrue}
\makeatother

\copyrightyear{2026}
\acmYear{2026}
\setcopyright{cc}
\setcctype{by}

\acmConference[MM '26] {Proceedings of the 34th ACM International Conference on Multimedia}{November 10--14, 2026}{Rio de Janeiro, Brazil.}
\acmBooktitle{Proceedings of the 34th ACM International Conference on Multimedia (MM '26), November 10--14, 2026, Rio de Janeiro, Brazil}
\acmDOI{10.1145/3767308.3835035}
\acmISBN{979-8-4007-2213-4/2026/11}

\begin{document}

\title[Training-Free Attention-Guided Switching Between Explicit and Latent Thoughts for MLLMs]{\emph{Balancing Efficiency and Efficacy}: \\ Training-Free Attention-Guided Switching \\ Between Explicit and Latent Thoughts for MLLMs}

\author{Haoqian Kang}
\orcid{0009-0004-8417-9956}
\authornotemark[1]
\affiliation{%
  \institution{Harbin Institute of Technology, Shenzhen}
  \city{Shenzhen}
  \country{China}
}
\email{26b951036@stu.hit.edu.cn}

\author{Liupeng Li}
\orcid{0000-0001-7741-9289}
\authornotemark[1]
\affiliation{%
  \institution{Harbin Institute of Technology, Shenzhen}
  \city{Shenzhen}
  \country{China}
}
\email{25b951045@stu.hit.edu.cn}

\author{Kuofeng Gao}
\orcid{0000-0002-5667-8238}
\affiliation{%
  \institution{Tsinghua Shenzhen International Graduate School, Tsinghua University}
  \city{Shenzhen}
  \country{China}
}
\email{gkf24@mails.tsinghua.edu.cn}

\author{Jinpeng Wang}
\orcid{0000-0002-4352-4897}
\authornotemark[2]
\affiliation{%
  \institution{Harbin Institute of Technology, Shenzhen}
  \city{Shenzhen}
  \country{China}
}
\email{wangjp26@gmail.com}

\author{Zhenyu Lu}
\orcid{0009-0009-2925-7030}
\affiliation{%
  \institution{Peng Cheng Laboratory}
  \city{Shenzhen}
  \country{China}
}
\affiliation{%
  \institution{Shenzhen Institutes of Advanced Technology, Chinese Academy of Sciences}
  \city{Shenzhen}
  \country{China}
}
\email{zhenyulu22@m.fudan.edu.cn}

\author{Bin Chen}
\orcid{0000-0002-4798-230X}
\affiliation{%
  \institution{Harbin Institute of Technology, Shenzhen}
  \city{Shenzhen}
  \country{China}
}
\email{chenbin2021@hit.edu.cn}

\author{Ke Chen}
\orcid{0000-0003-0928-5199}
\affiliation{%
  \institution{Peng Cheng Laboratory}
  \city{Shenzhen}
  \country{China}
}
\email{chenk02@pcl.ac.cn}

\author{Yaowei Wang}
\orcid{0000-0002-6110-4036}
\authornotemark[2]
\affiliation{%
  \institution{Harbin Institute of Technology, Shenzhen}
  \city{Shenzhen}
  \country{China}
}
\affiliation{%
  \institution{Peng Cheng Laboratory}
  \city{Shenzhen}
  \country{China}
}
\email{yaoweiwang@gmail.com}

\makeatletter
\g@addto@macro\@authornotes{%
  \footnotetext[1]{Equal Contribution.}
  \footnotetext[2]{Jinpeng Wang and Yaowei Wang are corresponding authors.}%
}
\makeatother

\renewcommand{\shortauthors}{Haoqian Kang et al.}

\definecolor{ForestGreen}{RGB}{34,139,34}

\newcommand{\dui}{\textcolor{teal}{\CheckmarkBold}}
\newcommand{\cuo}{\textcolor{purple}{\XSolidBrush}}
\newcommand{\ie}{\emph{i.e.},~}
\newcommand{\eg}{\emph{e.g.},~}
\newcommand{\wrt}{\emph{w.r.t.}~}
\newcommand{\aka}{\emph{aka}~}
\newcommand{\todo}{\color{red}{\textbf{TODO.}}~}

\newcommand{\xiaok}[1]{\left(#1\right)}
\newcommand{\zhongk}[1]{\left[#1\right]}
\newcommand{\dak}[1]{\left\{#1\right\}}
\newcommand{\jiaok}[1]{\left<#1\right>}
\newcommand{\shuk}[1]{\left\lVert#1\right\rVert}
\newcommand{\shuks}[1]{\left\lVert#1\right\rVert^2}
\newcommand{\shangk}[1]{\left\lceil #1 \right\rceil}
\newcommand{\xiak}[1]{\left\lfloor #1 \right\rfloor}

\newcommand{\argmax}[1]{{\mathop{\arg\mathrm{max}}_{#1}\,}}
\newcommand{\argmin}[1]{{\mathop{\arg\mathrm{min}}_{#1}\,}}
\newcommand{\pfrac}[2]{\frac{\partial #1}{\partial #2}}
\newcommand{\prob}[2]{p\xiaok{#1 \mid #2}}

\newcommand{\T}{\top}
\newcommand{\dif}{\mathop{}\!\mathrm{d}}
\newcommand{\biset}[1]{\{0,1\}^{#1}}
\newcommand{\ReLU}{\mathrm{ReLU}}
\newcommand{\relu}{\mathrm{ReLU}}
\newcommand{\sign}{\mathrm{sign}}
\newcommand\softmax{\mathrm{softmax}}
\newcommand\KL{D_{\mathrm{KL}}}
\newcommand\Var{\mathrm{Var}}
\newcommand\Cov{\mathrm{Cov}}
\newcommand{\Tr}{\mathrm{Tr}}
\newcommand{\tr}{\mathrm{tr}}
\newcommand{\dist}{\mathrm{dist}}
\newcommand{\concat}{\mathrm{concat}}
\newcommand{\mean}{\mathrm{mean}}
\newcommand{\diag}{\mathrm{diag}}
\newcommand{\cov}{\mathrm{cov}}

\newcommand{\range}[1]{{0,1,\cdots,#1}} 
\newcommand{\Range}[1]{{1,2,\cdots,#1}} 
\newcommand{\opseq}[3]{{#1_1 #3 #1_2 #3 \cdots #3 #1_{#2}}}
\newcommand{\seq}[2]{\opseq{#1}{#2}{,}}
\newcommand{\xseq}[2]{\opseq{#1}{#2}{\times}}

\newcommand{\bma}{\bm{a}}
\newcommand{\bmb}{\bm{b}}
\newcommand{\bmc}{\bm{c}}
\newcommand{\bmd}{\bm{d}}
\newcommand{\bme}{\bm{e}}
\newcommand{\bmf}{\bm{f}}
\newcommand{\bmg}{\bm{g}}
\newcommand{\bmh}{\bm{h}}
\newcommand{\bmi}{\bm{i}}
\newcommand{\bmj}{\bm{j}}
\newcommand{\bmk}{\bm{k}}
\newcommand{\bml}{\bm{l}}
\newcommand{\bmm}{\bm{m}}
\newcommand{\bmn}{\bm{n}}
\newcommand{\bmo}{\bm{o}}
\newcommand{\bmp}{\bm{p}}
\newcommand{\bmq}{\bm{q}}
\newcommand{\bmr}{\bm{r}}
\newcommand{\bms}{\bm{s}}
\newcommand{\bmt}{\bm{t}}
\newcommand{\bmu}{\bm{u}}
\newcommand{\bmv}{\bm{v}}
\newcommand{\bmw}{\bm{w}}
\newcommand{\bmx}{\bm{x}}
\newcommand{\bmy}{\bm{y}}
\newcommand{\bmz}{\bm{z}}
\newcommand{\bmzero}{\bm{0}}
\newcommand{\bmone}{\bm{1}}
\newcommand{\bmalpha}{\bm{\alpha}}
\newcommand{\bmbeta}{\bm{\beta}}
\newcommand{\bmgamma}{\bm{\gamma}}
\newcommand{\bmdelta}{\bm{\delta}}
\newcommand{\bmepsilon}{\bm{\epsilon}}
\newcommand{\bmtheta}{\bm{\theta}}
\newcommand{\bmiota}{\bm{\iota}}
\newcommand{\bmkappa}{\bm{\kappa}}
\newcommand{\bmlambda}{\bm{\lambda}}
\newcommand{\bmmu}{\bm{\mu}}
\newcommand{\bmnu}{\bm{\nu}}
\newcommand{\bmxi}{\bm{\xi}}
\newcommand{\bmpi}{\bm{\pi}}
\newcommand{\bmrho}{\bm{\rho}}
\newcommand{\bmsigma}{\bm{\sigma}}
\newcommand{\bmtau}{\bm{\tau}}
\newcommand{\bmupsilon}{\bm{\upsilon}}
\newcommand{\bmphi}{\bm{\phi}}
\newcommand{\bmchi}{\bm{\chi}}
\newcommand{\bmpsi}{\bm{\psi}}
\newcommand{\bmomega}{\bm{\omega}}
\newcommand{\bmA}{\bm{A}}
\newcommand{\bmB}{\bm{B}}
\newcommand{\bmC}{\bm{C}}
\newcommand{\bmD}{\bm{D}}
\newcommand{\bmE}{\bm{E}}
\newcommand{\bmF}{\bm{F}}
\newcommand{\bmG}{\bm{G}}
\newcommand{\bmH}{\bm{H}}
\newcommand{\bmI}{\bm{I}}
\newcommand{\bmJ}{\bm{J}}
\newcommand{\bmK}{\bm{K}}
\newcommand{\bmL}{\bm{L}}
\newcommand{\bmM}{\bm{M}}
\newcommand{\bmN}{\bm{N}}
\newcommand{\bmO}{\bm{O}}
\newcommand{\bmP}{\bm{P}}
\newcommand{\bmQ}{\bm{Q}}
\newcommand{\bmR}{\bm{R}}
\newcommand{\bmS}{\bm{S}}
\newcommand{\bmT}{\bm{T}}
\newcommand{\bmU}{\bm{U}}
\newcommand{\bmV}{\bm{V}}
\newcommand{\bmW}{\bm{W}}
\newcommand{\bmX}{\bm{X}}
\newcommand{\bmY}{\bm{Y}}
\newcommand{\bmZ}{\bm{Z}}
\newcommand{\bmGamma}{\bm{\Gamma}}
\newcommand{\bmDelta}{\bm{\Delta}}
\newcommand{\bmTheta}{\bm{\Theta}}
\newcommand{\bmLambda}{\bm{\Lambda}}
\newcommand{\bmXi}{\bm{\Xi}}
\newcommand{\bmPi}{\bm{\Pi}}
\newcommand{\bmSigma}{\bm{\Sigma}}
\newcommand{\bmUpsilon}{\bm{\Upsilon}}
\newcommand{\bmPhi}{\bm{\Phi}}
\newcommand{\bmPsi}{\bm{\Psi}}
\newcommand{\bmOmega}{\bm{\Omega}}

\newcommand{\calA}{\mathcal{A}}
\newcommand{\calB}{\mathcal{B}}
\newcommand{\calC}{\mathcal{C}}
\newcommand{\calD}{\mathcal{D}}
\newcommand{\calE}{\mathcal{E}}
\newcommand{\calF}{\mathcal{F}}
\newcommand{\calG}{\mathcal{G}}
\newcommand{\calH}{\mathcal{H}}
\newcommand{\calI}{\mathcal{I}}
\newcommand{\calJ}{\mathcal{J}}
\newcommand{\calK}{\mathcal{K}}
\newcommand{\calL}{\mathcal{L}}
\newcommand{\calM}{\mathcal{M}}
\newcommand{\calN}{\mathcal{N}}
\newcommand{\calO}{\mathcal{O}}
\newcommand{\calP}{\mathcal{P}}
\newcommand{\calQ}{\mathcal{Q}}
\newcommand{\calR}{\mathcal{R}}
\newcommand{\calS}{\mathcal{S}}
\newcommand{\calT}{\mathcal{T}}
\newcommand{\calU}{\mathcal{U}}
\newcommand{\calV}{\mathcal{V}}
\newcommand{\calW}{\mathcal{W}}
\newcommand{\calX}{\mathcal{X}}
\newcommand{\calY}{\mathcal{Y}}
\newcommand{\calZ}{\mathcal{Z}}

\newcommand{\bbC}{\mathbb{C}}
\newcommand{\bbE}{\mathbb{E}}
\newcommand{\bbI}{\mathbb{I}}
\newcommand{\bbN}{\mathbb{N}}
\newcommand{\bbQ}{\mathbb{Q}}
\newcommand{\bbR}{\mathbb{R}}
\newcommand{\bbZ}{\mathbb{Z}}

\newcommand{\tabincell}[2]{\begin{tabular}{@{}#1@{}}#2\end{tabular}}


\newcommand{\metricname}{\text{v2t-weight}\xspace}
\newcommand{\methodname}{\textsc{AGS}\xspace}

\begin{abstract}
Reasoning in Multimodal Large Language Models (MLLMs) requires both fine-grained visual perception and rigorous logical deduction. Explicit text-based Chain-of-Thought (CoT) is computationally expensive and prone to visual hallucinations, while existing latent reasoning methods typically require costly training. Furthermore, directly adapting training-free LLM reasoning mechanisms to the multimodal setting yields unstable performance. We identify that this failure stems from their reliance on token-level entropy, which fundamentally conflates perceptual ambiguity (e.g., unclear visual details) with logical uncertainty (e.g., complex reasoning steps). To overcome this bottleneck, we present a novel training-free inference strategy for MLLMs that explicitly decouples perception and reasoning. We propose a novel metric, the vision-to-text attention ratio, to dynamically gauge the model's cognitive focus. Guided by this metric, our proposed framework, \textbf{A}ttention-\textbf{G}uided \textbf{S}witching (\methodname), adaptively triggers latent reasoning for perceptual tokens to preserve high-fidelity visual information in the continuous space, while enforcing explicit text generation for logical tokens to maintain structural anchoring.
Extensive experiments demonstrate that our method achieves state-of-the-art performance, significantly improving both accuracy and inference efficiency by reducing autoregressive steps and latency. Code is released at \url{https://github.com/swordAndSnow/MM26-AGS}.
\end{abstract}

\ccsdesc[500]{Computing methodologies~Computer vision}
\ccsdesc[500]{Computing methodologies~Natural language processing}
\ccsdesc[300]{Computing methodologies~Knowledge representation and reasoning}

\keywords{multimodal large language models; latent reasoning; decoupling perception and reasoning}

\maketitle

\section{Introduction}
\label{sec: introduction}

\begin{figure}[t]
    \centering
    \includegraphics[width=1\linewidth]{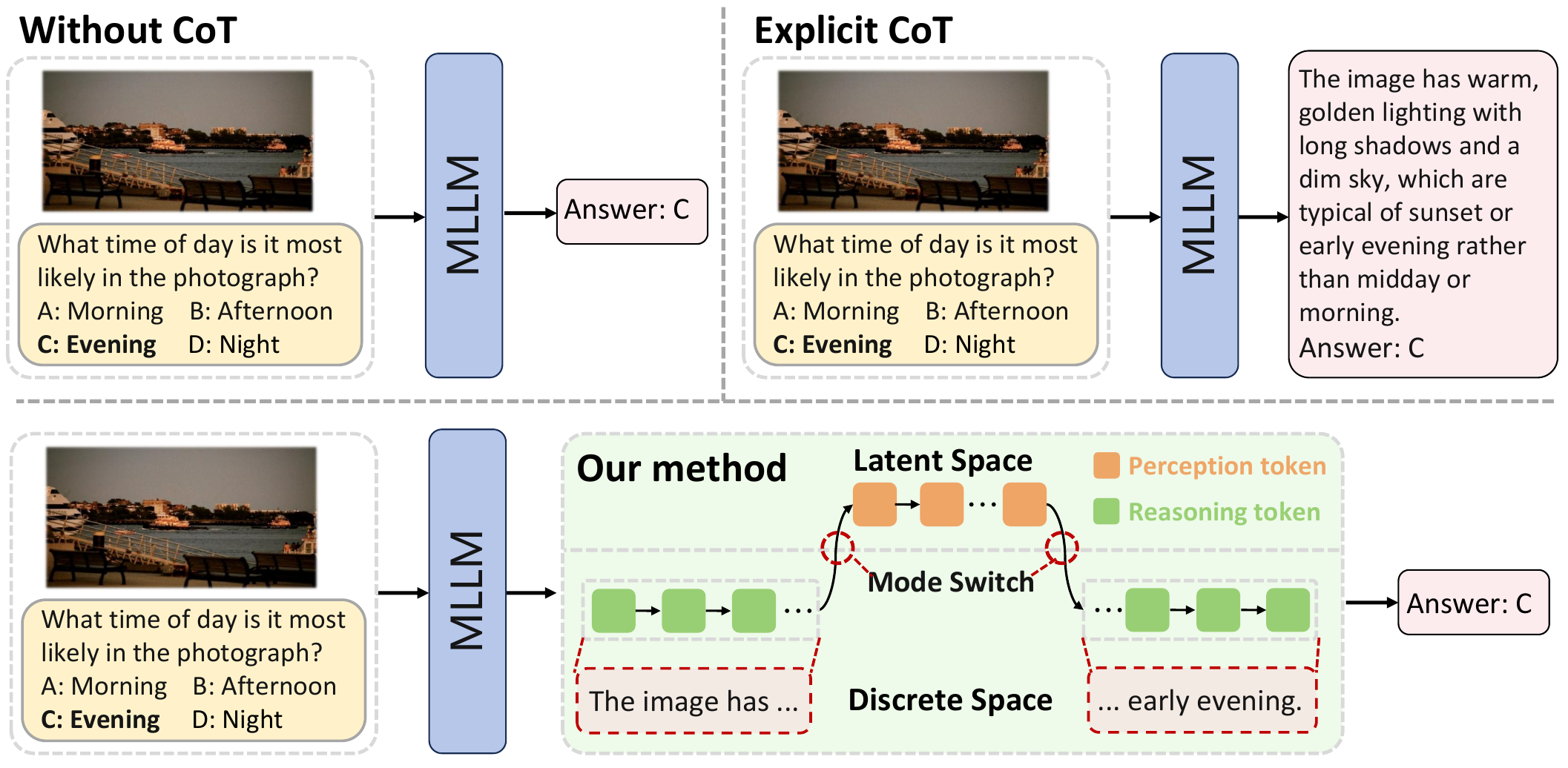}
    \caption{
Comparison of reasoning paradigms in MLLMs on a VQA example. \textbf{Without CoT}: directly predicts the answer. \textbf{Explicit CoT}: performs step-by-step reasoning in the discrete token space before producing the answer. \textbf{Our proposed method} (\methodname): interleaves perception and reasoning tokens during inference, dynamically routing visual processing to the latent space to bypass redundant discrete text generation.
    }
    \label{fig:cmp}
\end{figure}

Reasoning is a fundamental capability of Multimodal Large Language Models (MLLMs)~\cite{li2024llavaonevision, bai2025qwen3, wang2025internvl3, su2025thinking}, enabling complex cross-modal tasks such as scientific diagram analysis and mathematical problem-solving. Unlike pure text-based logic~\cite{guo2025deepseek, yeo2025demystifying}, multimodal reasoning inherently requires the simultaneous mastery of two distinct cognitive processes: perceiving fine-grained visual details and performing rigorous logical deduction.

The dominant reasoning paradigm in current MLLMs is explicit text-based chain-of-thought (CoT)~\cite{deng2025openvlthinker,huang2025vision-r1, yang2025r1-onevision}, which relies on the generation of discrete tokens during inference, as illustrated in Fig.~\ref{fig:cmp}. However, coercing complex visual concepts into discrete linguistic tokens often incurs significant computational inefficiency and exacerbates visual hallucinations. To mitigate these issues, recent works~\cite{li2025latentvisualReasoning, pham2025MCOUT, wang2025monet, yang2025machineimagery, zhang2025latentSketchpad} have explored latent reasoning, which interleaves text and image embeddings in a continuous space to emulate ``thinking with images"~\cite{zheng2025deepeyesTWI, zhang2025thyme}. Unfortunately, since such latent behaviors are not naturally aligned during standard pre-training, existing multimodal latent reasoning frameworks typically demand substantial training overhead through high-quality external datasets or strong teacher supervision.

To bypass the massive costs of retraining, a natural intuitive step is to adapt training-free latent reasoning frameworks originally designed for LLMs (e.g., SwiReasoning~\cite{shi2025swiR}) to the MLLM setting. These methods dynamically alternate between explicit text generation and implicit continuous embeddings based on token-level probability. However, our preliminary empirical evaluations reveal that such MLLM variants exhibit highly unstable performance, sometimes even underperforming vanilla explicit CoT. Guided by the principle of decoupling perception and reasoning~\cite{qiao2024prism, jia2025decoupling}, we diagnose a critical flaw in these prior arts: the reliance on token-level entropy as the sole metric for mode switching. In multimodal contexts, high output entropy conflates two fundamentally different challenges, perceptual ambiguity (e.g., struggling to identify a tiny visual object) and reasoning uncertainty (e.g., deliberating over a complex mathematical step). Treating these distinct cognitive bottlenecks uniformly leads to suboptimal routing decisions.

\begin{figure}[t]
    \centering
    \includegraphics[width=1\linewidth]{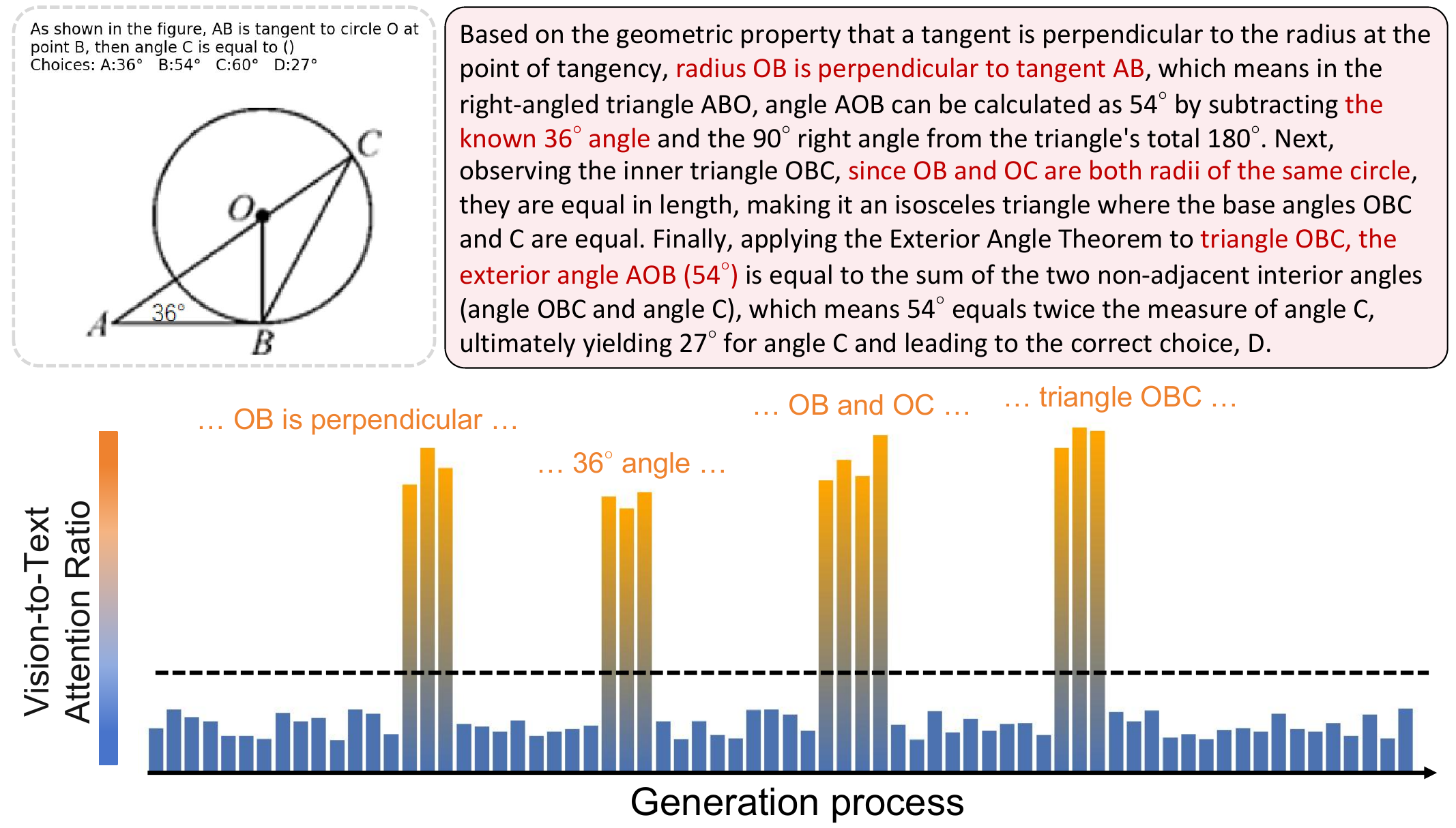}
    \caption{
Visualization of the proposed vision-to-text attention ratio ($R_A$) during generation on a geometric reasoning example. The bar chart demonstrates how attention ratio dynamically fluctuates throughout the inference phase. Notably, the distinct spikes in the ratio (orange bars) align precisely with the generation of perception-heavy visual concepts from the image, such as identifying spatial relationships and specific angles (highlighted in red in the text).
    }
    \label{fig:case}
\end{figure}

To address this entanglement, we argue that mode switching must dynamically adapt to whether the model is in a perception-dominant or logic-dominant phase during generation. Moving beyond the opaque nature of output probabilities, we delve into the model's internal representations and propose a novel, highly interpretable metric: the vision-to-text attention ratio. Concretely, the vision-to-text attention ratio is defined as the ratio of the attention allocated to visual tokens versus text tokens. This metric directly reflects the functional role of the current decoding step, accurately distinguishing perceptual tokens from reasoning tokens. As visualized in Fig.~\ref{fig:case}, tokens exhibiting higher vision-to-text attention ratios consistently align with perception-heavy concepts, empirically validating our hypothesis that this metric captures the true cognitive state of the model.

Empowered by the proposed metric, we introduce \textbf{A}ttention-\textbf{G}uided \textbf{S}witching (\methodname), a novel training-free inference strategy that performs asymmetric token routing. Since visual information is naturally dense and continuous, coercing it into textual tokens during perception-heavy phases inevitably causes information loss. Therefore, for perceptual tokens (where the model heavily attends to the image), \methodname adaptively triggers implicit latent reasoning. Specifically, it forms a probability-weighted mixture of token embeddings as inputs, enabling plug-and-play operation at inference time. This bypasses the linguistic bottleneck, allowing the model to process high-fidelity, continuous visual representations without forcing premature textualization that causes hallucinations. Conversely, since logical deduction relies on strict sequential dependencies, for reasoning tokens, the framework enables explicit text generation to provide strong logical anchoring and structural rigor, thereby minimizing deduction errors.

Extensive experiments demonstrate the superiority of \methodname. Compared to explicit text-based CoT and entropy-based SwiR-MLLM baselines, our method achieves state-of-the-art performance while remarkably accelerating the inference process.

Our contributions are summarized as follows:
\setlist{nolistsep}
\begin{itemize}[leftmargin=1.5em]
\item We identify entropy entanglement in multimodal reasoning,
where token-level entropy conflates perceptual ambiguity with
logical uncertainty, resulting in unstable mode switching.
\item We propose the vision-to-text attention ratio, an interpretable
token-level metric to distinguish perception-dominant and
logic-dominant decoding steps.
\item Based on this metric, we develop \methodname, a training-free strategy that adaptively routes perceptual tokens to the continuous latent space and logical tokens to the explicit text space.
\item Extensive evaluations of challenging multimodal benchmarks validate that our method achieves superior accuracy while significantly reducing autoregressive decoding steps and overall inference latency.
\end{itemize}

\section{Related Works}
\label{sec:related_work}

\subsection{MLLM Reasoning}
Recent advances in Multimodal Large Language Models
(MLLMs)~\cite{tong2026swimbird, chen2025sft-RL,
meng2025mm-eureka, fan2025sophiavl, luo2025ursa,
lai2025mini, yang2025drim, lu2026affinspace} have extended visual
question answering toward complex multimodal reasoning. Early
methods~\cite{liu2023llava, li2024llavaonevision} mainly follow the
``thinking about images'' paradigm, where visual inputs are encoded
once and statically consumed by the LLM. This static design may weaken
visual grounding over long reasoning chains and increase hallucinations.

Recent works~\cite{shao2024vcot, zhao2025uv-cot, gao2025icot}
instead pursue ``thinking with images,'' treating visual inputs as an
interactive reasoning workspace. Representative approaches include
tool-augmented visual search~\cite{wu2025grounded-cot,
cheng2025visual-thoughts, li2026cvsearch}, programmatic visual
manipulation~\cite{hu2024visual-sketchpad}, interpretable
reasoning--perception alignment~\cite{lu2026segcompass}, and visual
imagination~\cite{zhang2025latentSketchpad}. These approaches improve
fidelity but often require extra tools, computation, or training.
Broader efficiency studies compress visual tokens~\cite{zhou2026comptrack}
or replace attention in vision backbones~\cite{scribano2026accelerating},
yet do not optimize MLLM reasoning trajectories. Long outputs further
increase MLLM inference cost~\cite{gao2024inducing,
wang2025vlminferslow}. In contrast, our training-free method dynamically
switches between explicit and latent reasoning, reducing generation
steps without modifying model parameters.

\subsection{Decoupling Perception and Reasoning}
Coupling visual perception with symbolic reasoning can undermine
reasoning stability in complex multimodal tasks. Early methods adopt
macro-level decoupling through isolated modules, such as LLM--MLLM
collaboration~\cite{jia2025decoupling} or two-stage
pipelines~\cite{qiao2024prism, gou2025RAPID}. Although these
designs can reduce hallucinations, they incur additional latency and
cannot fully capture the interleaved nature of multimodal cognition.

Recent approaches~\cite{li2026rethinking, huang2025spotlight,
lu2026bridging} instead perform token-level decoupling within a
single autoregressive trajectory, distinguishing perceptual and logical
tokens through signals such as visual sensitivity, distributional
shifts, or hidden-state similarity. However, they still require
additional supervision, curated data, or training overhead. Our method
instead uses a training-free white-box metric based on cross-modal
attention ratios to dynamically distinguish and switch between
perception-dominant and logic-dominant states during inference.

\subsection{Latent Reasoning}
Explicit Chain-of-Thought (CoT)~\cite{wang2025multimodal-cot}
improves reasoning but incurs substantial latency by autoregressively
generating intermediate text. Latent reasoning~\cite{zhang2025softthinking,
tan2025think-silent, hao2024training-latent} reduces this cost by
performing intermediate computation in continuous representation
spaces. Originating in LLMs~\cite{xu2025softcot, xu2025softcot++,
zhang2025softthinking}, existing methods typically propagate latent
reasoning through previous hidden states, probability-weighted
vocabulary embeddings, or dedicated latent tokens.

Latent reasoning has recently been extended to
MLLMs~\cite{li2025latentvisualReasoning, pham2025MCOUT,
wang2025monet, yang2025machineimagery, zhang2025latentSketchpad}
to support ``thinking with images.'' However, existing multimodal
methods generally rely on resource-intensive retraining and
fine-grained supervision to align latent trajectories. Directly applying
training-free LLM routing methods, such as entropy-based
SwiReasoning~\cite{shi2025swiR}, is also suboptimal because output
entropy conflates perceptual ambiguity with logical uncertainty. Our
framework addresses this gap with an attention-guided, asymmetric,
and training-free latent reasoning strategy for MLLMs.

\begin{figure*}[t]
    \centering
    \includegraphics[width=\textwidth]{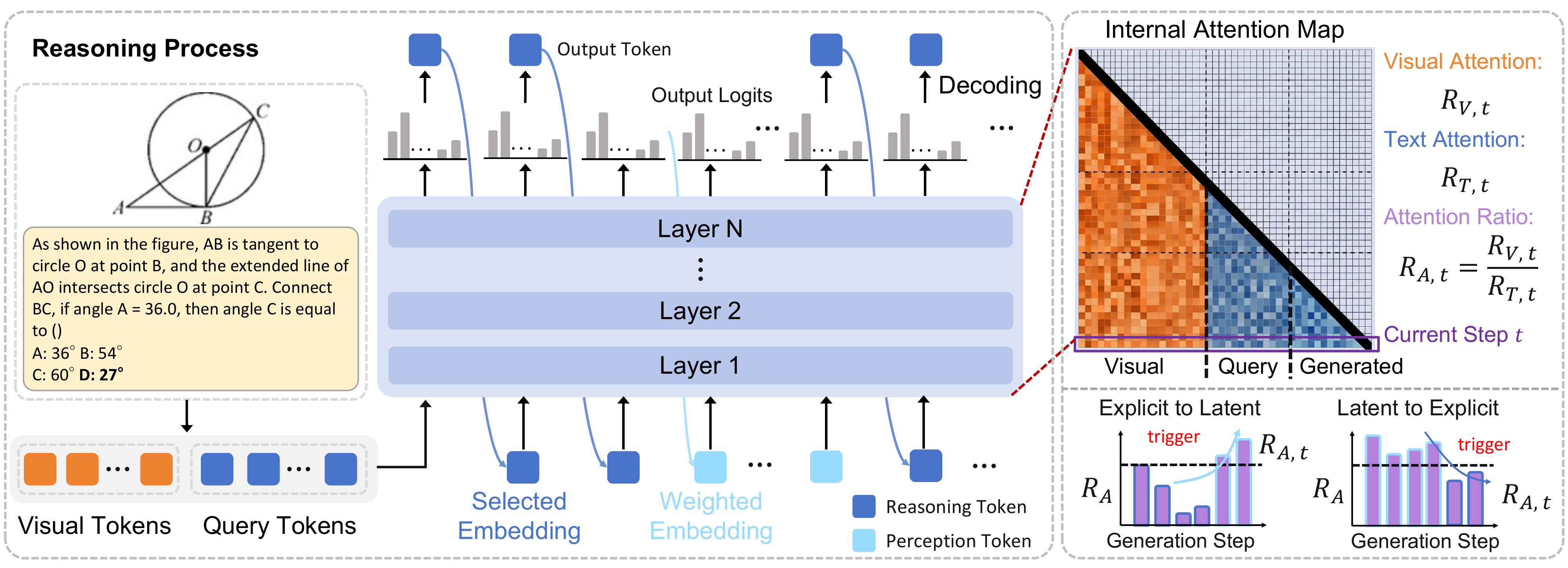}
    \caption{Illustration of \methodname. The overall inference process (left) shows the model dynamically shifting its reasoning mode based on the calculated vision-to-text attention ratio, $R_{A,t}$. High $R_{A,t}$ value triggers a switch from explicit logic to latent perception, while low $R_{A,t}$ value causes a transition from latent perception back to explicit logic, as visualized in the bottom-right plots. The specific calculation for the $R_{A,t}$ is detailed in the top-right panel.}
    \label{fig:method}
\end{figure*}

\section{Method}
\label{sec:method}

\textbf{Overview.} In this section, we present the \methodname framework. We first provide preliminaries on standard and latent MLLM reasoning formulations in Section~\ref{subsec:preliminary}. Next, in Section~\ref{subsec:v2t-weight}, we introduce the vision-to-text attention ratio to quantitatively decouple perceptual and logical tokens. Finally, Section~\ref{subsec:SwiPR} details the token-level dynamic routing mechanism that seamlessly switches between latent and explicit reasoning modes.

\subsection{Preliminary: MLLM Reasoning}
\label{subsec:preliminary}

\paragraph{Vision and Language Inputs} Let $\mathbf{I}$ and $Q$ denote the image and question fed into a Multimodal Large Language Model (MLLM). The image $\mathbf{I}$ is first processed by a vision encoder to extract visual features, which are subsequently mapped into the LLM embedding space $\mathbb{R}^d$ via a cross-modal projector. This yields a sequence of $N$ visual embeddings, denoted as $\mathbf{E}_v = [\mathbf{e}_{v,1}, \mathbf{e}_{v,2}, \dots, \mathbf{e}_{v,N}] \in \mathbb{R}^{N \times d}$, where each $\mathbf{e}_{v,i} \in \mathbb{R}^d$. Concurrently, the textual input $Q$ is tokenized and embedded into a sequence of $M$ textual embeddings $\mathbf{E}_t = [\mathbf{e}_{t,1}, \mathbf{e}_{t,2}, \dots, \mathbf{e}_{t,M}] \in \mathbb{R}^{M \times d}$. Finally, the visual and textual embeddings are concatenated along the sequence dimension to form a unified joint representation $\mathbf{E} = [\mathbf{E}_v; \mathbf{E}_t] = [\mathbf{e}_1, \mathbf{e}_2, \dots, \mathbf{e}_L] \in \mathbb{R}^{L \times d}$, where $L = N + M$. This consolidated sequence $\mathbf{E}$ serves as the foundational input for subsequent reasoning within the pre-trained LLM.

\paragraph{Autoregressive Forward Propagation} The reasoning backbone of the MLLM is a pre-trained LLM, parameterized by $\theta$ and denoted as $\mathcal{M}_{\theta}$. In standard autoregressive generation, the previously generated sequence is maintained entirely as a sequence of continuous embeddings, denoted as $\mathbf{E}_{<t} = [\mathbf{e}_1, \mathbf{e}_2, \dots, \mathbf{e}_{t-1}] \in \mathbb{R}^{(t-1) \times d}$. At each generation step $t$, conditioned on the initial multimodal context $\mathbf{E}$ and the generated sequence $\mathbf{E}_{<t}$, the model predicts the probability distribution of the t-th token:
\begin{equation}
\mathbf{p}_t = \mathcal{M}_\theta \big(\mathbf{E}, \mathbf{E}_{<t}\big) \in \Delta^{|\mathcal{V}|-1},
\label{eq:mlrm-forward-prob}
\end{equation}
where $\mathcal{V}$ is the vocabulary of the LLM, and $\Delta^{|\mathcal{V}|-1}$ represents the corresponding probability simplex. Based on this distribution, a hard discrete token $y_t \in \mathcal{V}$ is determined via a decoding strategy $\mathcal{D}$ (e.g., greedy search or nucleus sampling):
\begin{equation}
y_t = \mathcal{D}(\mathbf{p}_t).
\label{eq:mlrm-forward-decode}
\end{equation}
Subsequently, this discrete token $y_t$ is mapped back into the continuous representation space via the model's vocabulary embedding function $\text{Emb}: \mathcal{V} \rightarrow \mathbb{R}^d$. The continuous embedding for the current step is obtained as:
\begin{equation}
\mathbf{e}_t = \text{Emb}(y_t).
\label{eq:mlrm-forward-lookup}
\end{equation}
Finally, this selected embedding $\mathbf{e}_t$ is appended to the continuous sequence, forming $\mathbf{E}_{<t+1} = [\mathbf{E}_{<t}; \mathbf{e}_t]$, which acts as the input state for the next autoregressive step $t+1$.

\paragraph{Latent Reasoning Forward} Unlike standard autoregressive generation that relies on discrete token decoding, latent reasoning propagates information entirely within the continuous embedding space. Instead of sampling a hard discrete token $y_t$ and executing a discrete look-up ($\mathbf{e}_t = \text{Emb}(y_t)$), this approach utilizes the predicted probability distribution $\mathbf{p}_t$ to directly compute an expected, ``soft'' token embedding. Let $\mathbf{W} \in \mathbb{R}^{|\mathcal{V}| \times d}$ denote the vocabulary weight matrix corresponding to the embedding function $\text{Emb}(\cdot)$. The continuous latent representation $\tilde{\mathbf{e}}_t \in \mathbb{R}^d$ for the current step is computed as the probability-weighted sum of all vocabulary embeddings:
\begin{equation}
\tilde{\mathbf{e}}_t = \mathbf{W}^\top \mathbf{p}_t.
\label{eq:latent-forward}
\end{equation}
This continuous embedding $\tilde{\mathbf{e}}_t$ is then appended to the sequence in place of $\mathbf{e}_t$, forming the updated continuous context $\tilde{\mathbf{E}}_{<t+1} = [\mathbf{E}_{<t}; \tilde{\mathbf{e}}_t]$ for the subsequent reasoning step $t+1$. This fully continuous formulation effectively bypasses the information bottleneck of discrete token selection, allowing the aggregated embedding to carry and propagate richer distributional semantics within a single forward pass.

\subsection{Decoupling Perception and Reasoning}
\label{subsec:v2t-weight}

As discussed earlier, tokens play distinct functional roles during inference. To quantitatively capture this, we propose a novel metric, the attention ratio $R_{A}$, to characterize the specific role of each token at any given generation step.

Specifically, let $N_l$ and $N_h$ denote the total number of attention layers and attention heads within the LLM, respectively. Let $\mathcal{I}_v$ and $\mathcal{I}_t$ represent the index sets of the visual and textual tokens in the current context. As the autoregressive generation proceeds, the size of the visual index set $|\mathcal{I}_v|$ remains constant, whereas the textual index set $\mathcal{I}_t$ dynamically expands. Let $\alpha_{t,j}^{(l,h)}$ denote the attention weight assigned by the current query token at step $t$ to the key token $j$, computed by the $h$-th attention head in the $l$-th layer.

To mitigate the bias introduced by the inherent length disparity between visual and textual contexts, we first calculate the average attention allocated at the current generation step $t$ to visual tokens, denoted as $A_{V,t}$, and textual tokens, denoted as $A_{T,t}$. These are averaged across all layers and heads:
\begin{equation}
\begin{aligned}
A_{V,t} &= \frac{1}{|\mathcal{I}_v| N_l N_h} \sum_{l=1}^{N_l} \sum_{h=1}^{N_h} \sum_{j \in \mathcal{I}_v} \alpha_{t,j}^{(l,h)}, \\
A_{T,t} &= \frac{1}{|\mathcal{I}_t| N_l N_h} \sum_{l=1}^{N_l} \sum_{h=1}^{N_h} \sum_{k \in \mathcal{I}_t} \alpha_{t,k}^{(l,h)}.
\end{aligned}
\label{eq:avg_attention}
\end{equation}
Based on these normalized attention values, we define the overall attention ratio at step $t$ as:
\begin{equation}
R_{A,t} = \frac{A_{V,t}}{A_{T,t}}.
\label{eq:metric}
\end{equation}

As illustrated in Figure~\ref{fig:method} (top-right), $R_{A,t}$ quantifies the dynamic information-seeking preference of the model during generation. A high $R_{A,t}$ indicates a dominant attention allocation toward the visual tokens, signifying that the model is actively extracting visual information. We define such tokens as serving a perception role. Conversely, a low $R_{A,t}$ reflects a structural shift in reliance toward the textual context for semantic deduction and logical progression. Accordingly, we characterize these tokens as functioning in a reasoning role.

\subsection{Reasoning Mode Switching}
\label{subsec:SwiPR}

Building upon the token role decoupling, \methodname dynamically
routes each generation step to either the latent or explicit
reasoning mode. Since the absolute magnitude of visual attention
can vary across samples and reasoning phases due to differences in
image complexity and multimodal context, a single global threshold
may not generalize well. We therefore adopt a phase-wise dynamic
threshold, using a phase-specific reference to capture relative
attention changes within each reasoning phase.

Specifically, let $t_k$ denote the step at which the most recent mode switch occurred. The threshold for the current reasoning phase, denoted as $\tau_k$, is dynamically anchored to the attention ratio of the very first token generated in this phase, i.e., $\tau_k = R_{A, t_k}$. Then, when $R_{A,t} \ge \tau_k$, the model focuses on visual perception and operates in the latent reasoning mode to propagate information smoothly within the continuous embedding space. Conversely, when $R_{A,t} < \tau_k$, it indicates a structural shift towards textual reliance, prompting the model to switch to the explicit reasoning mode to produce discrete semantic tokens.

To ensure generation quality and logical coherence, we impose two critical constraints on the switching mechanism:
\textbf{Minimum Maintenance Window ($W$):} To prevent fragmented logical chains caused by overly frequent mode oscillation, we enforce a minimum duration for the explicit reasoning mode. If the model switches to explicit reasoning at step $t_{\text{exp}}$, it must remain in this state for at least $W$ consecutive steps, temporarily overriding the dynamic $R_{A,t}$ condition.
\textbf{Maximum Switch Budget ($C$):} Unconstrained reasoning loops may lead to non-termination. We track the cumulative number of mode transitions, denoted as $c_t$. Once $c_t$ reaches a predefined maximum budget $C$, we force the termination of the thinking process by appending the \texttt{</think>} token. Thereafter, the model exclusively adopts standard explicit reasoning until the generation concludes.

Formally, let $S_t \in \{0, 1\}$ denote the reasoning state at step $t$, where $0$ indicates latent reasoning and $1$ indicates explicit reasoning. Let $c_{t-1}$ denote the cumulative number of mode transitions prior to step $t$. Accounting for the dynamic threshold, the window constraint, and the maximum switch budget, the state indicator is determined by:
\begin{equation}
S_t = \mathbb{I}\big(R_{A,t} < \tau_k \lor (S_{t-1} = 1 \land t - t_{\text{exp}} < W) \lor c_{t-1} \ge C\big),
\end{equation}
where $\mathbb{I}(\cdot)$ is the indicator function. Following this state determination, the transition tracker is updated as $c_t = c_{t-1} + \mathbb{I}(S_t \neq S_{t-1})$. Let $\mathbf{e}_t^*$ denote the final selected embedding to be appended to the sequence for the subsequent step. To strictly enforce the stopping mechanism without causing an infinite loop, we inject the termination identifier exactly once at the step the budget is reached. The complete forward routing is formulated as:
\begin{equation}
\mathbf{e}_t^* =
\begin{cases}
\mathbf{e}_{\text{stop}}, & \text{if } c_t = C \text{ and } c_{t-1} < C \\
\tilde{\mathbf{e}}_t, & \text{if } c_t < C \text{ and } S_t = 0 \\
\text{Emb}(\hat{y}_t), & \text{otherwise}
\end{cases},
\label{eq:switching}
\end{equation}
where $\mathbf{e}_{\text{stop}} \in \mathbb{R}^d$ is the predefined continuous embedding of the termination identifier (e.g., \texttt{</think>}), $\tilde{\mathbf{e}}_t$ is the probability-weighted soft embedding from Eq.~\ref{eq:latent-forward}, and $\text{Emb}(\hat{y}_t)$ is the standard discrete embedding anchored by the greedy decoding token $\hat{y}_t = \mathop{\arg\max} \mathbf{p}_t$. This routing design empowers the MLLM to autonomously allocate computational bandwidth between continuous visual perception and discrete logical deduction, while mathematically guaranteeing deterministic termination.

\section{Experiments}
\label{sec:experiments}

\begin{table*}[t]
\captionsetup{skip=3pt}
\centering
\scriptsize
\setlength{\tabcolsep}{3.5pt} 
\caption{Quantitative evaluation across diverse multimodal reasoning tasks. We report both Accuracy (\textbf{Acc}, \%) and Efficiency (\textbf{Eff}), measured by the average number of autoregressive generation steps. Best results are highlighted in bold.}
\label{tab:main_results}
\makebox[\textwidth][c]{
\resizebox{\textwidth}{!}{
\begin{tabular}{l cccccccccccccc}
\toprule

\multirow{3}{*}[-0.6em]{Model \& Method} & \multicolumn{8}{c}{Math Reasoning} & \multicolumn{4}{c}{STEM and General} & \multicolumn{2}{c}{\multirow{2}{*}[-0.2em]{\textbf{Avg}}} \\
\cmidrule(lr){2-9} \cmidrule(lr){10-13}
& \multicolumn{2}{c}{MathVerse} & \multicolumn{2}{c}{MathVision} & \multicolumn{2}{c}{MathVista} & \multicolumn{2}{c}{WeMath} & \multicolumn{2}{c}{M$3$CoT} & \multicolumn{2}{c}{ScienceQA} & \multicolumn{2}{c}{} \\
\cmidrule(lr){2-3} \cmidrule(lr){4-5} \cmidrule(lr){6-7} \cmidrule(lr){8-9} \cmidrule(lr){10-11} \cmidrule(lr){12-13} \cmidrule(lr){14-15}
& Acc $\uparrow$ & Eff $\downarrow$ & Acc $\uparrow$ & Eff $\downarrow$ & Acc $\uparrow$ & Eff $\downarrow$ & Acc $\uparrow$ & Eff $\downarrow$ & Acc $\uparrow$ & Eff $\downarrow$ & Acc $\uparrow$ & Eff $\downarrow$ & \textbf{Acc} $\uparrow$ & \textbf{Eff} $\downarrow$ \\
\midrule

\multicolumn{15}{l}{\textbf{InternVL3.5 Series}}\\ 
\midrule
InternVL3.5-2B  & 35.6 & 365 & 9.4 & 494 & 23.2 & 253 & 51.6 & 372 & 50.6 & 288 & 77.1 & 193 & 41.3 & 328 \\
\rowcolor{blue!10}
InternVL3.5-2B \textit{(+ \methodname{})} & 36.9 & 362 & 10.3 & 490 & 23.0 & 236 & 53.2 & 348 & 52.8 & 271 & 81.8 & 176 & 43.0 & 314 \\
\addlinespace 
InternVL3.5-4B  & 47.8 & 341 & 15.9 & 486 & 36.0 & 260 & 65.8 & 286 & 64.8 & 268 & 88.4 & 158 & 53.1 & 300 \\
\rowcolor{blue!10}
InternVL3.5-4B \textit{(+ \methodname{})} & 50.0 & 337 & 16.9 & 472 & 35.7 & 257 & 65.9 & 274 & 66.0 & 272 & 87.9 & 144 & 53.7 & 293 \\
\addlinespace 
InternVL3.5-8B  & 48.3 & 315 & 18.8 & 671 & 39.0 & 251 & 70.8 & 341 & 67.0 & 207 & 90.3 & 133 & 55.7 & 320 \\
\rowcolor{blue!10}
InternVL3.5-8B \textit{(+ \methodname{})} & 48.8 & 302 & 18.2 & 434 & 38.5 & 219 & 70.9 & 263 & \textbf{68.0} & 209 & \textbf{90.5} & 121 & 55.8 & 258 \\
\midrule 

\multicolumn{15}{l}{\textbf{Qwen3-VL Series}}\\
\midrule
Qwen3-VL-2B-Thinking  & 34.1 & 3214 & 5.8 & 3941 & 27.2 & 2156 & 32.4 & 3364 & 48.6 & 2434 & 69.3 & 1413 & 36.2 & 2754 \\
\rowcolor{blue!10} 
Qwen3-VL-2B-Thinking \textit{(+ \methodname{})} & 42.6 & 1481 & 9.6 & 1970 & 37.1 & 1112 & 48.6 & 1512 & 54.2 & 1225 & 73.8 & 800 & 44.3 & 1350 \\
\addlinespace 
Qwen3-VL-4B-Thinking  & 44.1 & 3136 & 11.1 & 3868 & 33.3 & 2092 & 43.1 & 3217 & 57.5 & 2255 & 81.3 & 1242 & 45.1 & 2635 \\
\rowcolor{blue!10}
Qwen3-VL-4B-Thinking \textit{(+ \methodname{})} & 60.0 & 1144 & 17.7 & 1602 & \textbf{39.6} & 915 & 69.4 & 1114 & 65.8 & 985 & 88.2 & 643 & 56.8 & 1067 \\
\addlinespace 
Qwen3-VL-8B-Thinking  & 55.9 & 2574 & 10.7 & 3693 & 30.5 & 1746 & 62.9 & 2489 & 62.7 & 1881 & 87.5 & 898 & 51.7 & 2214 \\
\rowcolor{blue!10}
Qwen3-VL-8B-Thinking \textit{(+ \methodname{})} & \textbf{64.7} & 1002 & \textbf{18.9} & 1599 & 35.3 & 861 & \textbf{76.8} & 867 & 67.4 & 861 & 89.8 & 529 & \textbf{58.8} & 953 \\

\bottomrule
\end{tabular}
}}
\end{table*}
\begin{table}[h]
  \centering
  \setlength{\tabcolsep}{3pt} 
  \caption{Evaluation on the POPE Benchmark.}
  \vspace{-1em}
  \label{tab:pope_hallucination_sidebyside}
  \makebox[\columnwidth][c]{
  \resizebox{\columnwidth}{!}{
    \begin{tabular}{lcccc | lcccc}
    \toprule
    \multirow{2}{*}{\textbf{\textit{InternVL3.5}}} & \multicolumn{2}{c}{\textbf{Base}} & \multicolumn{2}{c|}{+ \textbf{AGS}} & 
    \multirow{2}{*}{\textbf{\textit{Qwen3-VL}}} & \multicolumn{2}{c}{\textbf{Base}} & \multicolumn{2}{c}{+ \textbf{AGS}} \\ 
    \cmidrule(lr){2-3} \cmidrule(lr){4-5} \cmidrule(lr){7-8} \cmidrule(l){9-10} 
     & \textbf{Acc $\uparrow$} & \textbf{Eff $\downarrow$} & \textbf{Acc $\uparrow$} & \textbf{Eff $\downarrow$} & 
     & \textbf{Acc $\uparrow$} & \textbf{Eff $\downarrow$} & \textbf{Acc $\uparrow$} & \textbf{Eff $\downarrow$} \\ 
    \midrule
    ~~2B & 83.6 & 78 & \cellcolor{blue!10}\textbf{84.5} & \cellcolor{blue!10}79 & 
    ~~2B-Think & 86.8 & 112 & \cellcolor{blue!10}\textbf{87.2} & \cellcolor{blue!10}104 \\
    ~~4B & 81.6 & 74 & \cellcolor{blue!10}\textbf{82.2} & \cellcolor{blue!10}71 & 
    ~~4B-Think & 87.0 & 130 & \cellcolor{blue!10}\textbf{87.1} & \cellcolor{blue!10}127 \\
    ~~8B & 82.8 & 79 & \cellcolor{blue!10}\textbf{83.9} & \cellcolor{blue!10}76 & 
    ~~8B-Think & 86.7 & 93  & \cellcolor{blue!10}\textbf{86.8} & \cellcolor{blue!10}88 \\ 
    \bottomrule
    \end{tabular}
  }}
  \vspace{-1.1em}
\end{table}

\subsection{Experimental Setup}
\paragraph{Models}
We evaluate \methodname on two state-of-the-art MLLM families:
Qwen3-VL-Thinking~\cite{bai2025qwen3} and
InternVL3.5~\cite{wang2025internvl3}, each covering 2B, 4B, and
8B variants. Qwen3-VL-Thinking typically produces long
deliberative chains, whereas InternVL3.5 follows a substantially
more concise reasoning pattern. Their distinct architectures,
reasoning behaviors, and parameter scales allow us to assess
whether \methodname generalizes across both computation-intensive
and compact inference regimes, rather than relying on a particular
model family or chain-length distribution.

\paragraph{Evaluation Benchmarks}
We evaluate our method on six challenging benchmarks spanning
three multimodal reasoning domains:
\begin{itemize}[leftmargin=1.5em, itemsep=2pt, topsep=2pt]
    \item \textbf{Mathematical Reasoning:}
    \textit{MathVista}~\cite{lu2023mathvista} evaluates
    multi-skill visual mathematics,
    \textit{MathVision}~\cite{wang2024mathvision} contains
    competition-level problems,
    \textit{MathVerse}~\cite{zhang2024mathverse} reduces textual
    bias to assess visual reasoning, and
    \textit{WeMath}~\cite{qiao2025wemath} focuses on multi-step
    mathematical reasoning.

    \item \textbf{STEM Reasoning:}
    \textit{ScienceQA}~\cite{lu2022scienceQA} evaluates
    multimodal scientific reasoning with background knowledge.

    \item \textbf{General Multimodal Reasoning:}
    \textit{M$^3$CoT}~\cite{chen2024m3cot} evaluates general
    multimodal multi-step reasoning across diverse scenarios.
\end{itemize}
For the additional baseline comparison, we also include
V$^\ast$~\cite{wu2024v}, which evaluates fine-grained visual
perception.

\paragraph{Baselines}
We primarily compare \methodname with \textbf{Explicit CoT},
which generates all intermediate reasoning steps as discrete text
before producing the final answer. To evaluate the proposed routing
metric, we additionally compare it with the entropy-based strategy
used in SwiReasoning~\cite{shi2025swiR} in the ablation study.
For multimodal latent reasoning, we compare with
\textbf{LEAD}~\cite{xu2026thinking}. Both LEAD and \methodname
use the same Qwen3-VL-4B-Thinking checkpoint, enabling a
checkpoint-controlled comparison that more directly isolates the
effect of the latent reasoning strategy.

\paragraph{Metrics}
We report \textbf{Task Accuracy} for answer correctness and \textbf{Inference Efficiency} as the average number of autoregressive generation steps. Unlike visible output-token counts, generation steps also account for latent reasoning steps, which perform forward propagation in the continuous embedding space without emitting discrete tokens. This metric therefore provides a fairer estimate of the actual inference cost.

\paragraph{Implementation Details}
We implement \methodname in PyTorch and evaluate it on NVIDIA RTX A6000 GPUs, following the official HuggingFace inference pipelines for Qwen3-VL and InternVL3.5. The maximum sequence length is 4096. We set the minimum explicit-reasoning maintenance window to $W=512$ and the maximum switch budget to $C=4$ for all experiments. Explicit tokens are generated using multinomial sampling with \texttt{do\_sample=True} and a temperature of $0.6$.

\subsection{Main Result}
Table~\ref{tab:main_results} reports results across two MLLM
families and six reasoning benchmarks. Overall, \methodname
improves both accuracy and inference efficiency, alleviating the
accuracy--efficiency trade-off of Explicit CoT.
Table~\ref{tab:pope_hallucination_sidebyside} further shows
consistent POPE improvements, indicating reduced object
hallucinations without sacrificing efficiency.

The gains are particularly pronounced for Qwen3-VL-Thinking,
which typically produces long reasoning chains. On the 8B model,
\methodname reduces the average generation cost from $2214$ to
$953$ steps ($\sim57\%$) while improving accuracy from $51.7\%$
to $58.8\%$. The 2B and 4B variants similarly reduce generation
steps by approximately $50\%$ and $60\%$, with absolute accuracy
gains of $8.1$ and $11.7$ percentage points, respectively.

The improvements are especially strong on visual-mathematical
tasks. On WeMath, \methodname raises Qwen3-VL-4B-Thinking
accuracy from $43.1\%$ to $69.4\%$; on MathVerse, the 4B and
8B models gain $15.9$ and $8.8$ percentage points. These results
suggest that latent routing avoids redundant textualization while
preserving visual information.

\begin{table*}[t] 
\captionsetup{skip=3pt}
\centering
\small 
\setlength{\tabcolsep}{7pt} 
\caption{Ablation study on the reasoning mode transition metric using the Qwen3-VL-8B-Thinking model on MathVerse, WeMath, M$3$CoT, and ScienceQA.}
\label{tab:ablation_metric}
\makebox[\textwidth][c]{
\begin{tabular}{l cccccccccc} 
\toprule
\multirow{2}{*}[-0.2em]{Method / Routing Metric} & \multicolumn{2}{c}{MathVerse} & \multicolumn{2}{c}{WeMath} & \multicolumn{2}{c}{M$3$CoT} & \multicolumn{2}{c}{ScienceQA} & \multicolumn{2}{c}{\textbf{Avg}} \\
\cmidrule(lr){2-3} \cmidrule(lr){4-5} \cmidrule(lr){6-7} \cmidrule(lr){8-9} \cmidrule(lr){10-11}
& Acc $\uparrow$ & Eff $\downarrow$ & Acc $\uparrow$ & Eff $\downarrow$ & Acc $\uparrow$ & Eff $\downarrow$ & Acc $\uparrow$ & Eff $\downarrow$ & \textbf{Acc} $\uparrow$ & \textbf{Eff} $\downarrow$ \\
\midrule
Explicit CoT \textit{(No Routing)} & 55.9 & 2574 & 62.9 & 2489 & 62.7 & 1881 & 87.5 & 898 & 67.3 & 1961 \\
Token Entropy & 58.8 & 1068 & 68.9 & 1001 & 66.4 & 850 & 88.8 & 535 & 70.7 & 864 \\
\rowcolor{blue!10}
Attention Ratio (Ours) & \textbf{64.7} & 1002 & \textbf{76.8} & 867 & \textbf{67.4} & 861 & \textbf{89.8} & 529 & \textbf{74.7} & 815 \\
\bottomrule
\end{tabular}%
}
\end{table*}
\methodname also generalizes to the more concise InternVL3.5
family, whose baselines average about $300$ generation steps. For
InternVL3.5-8B, it reduces the cost from $320$ to $258$ steps
while maintaining comparable average accuracy ($55.7\%$ versus
$55.8\%$) and achieving $90.5\%$ on ScienceQA. Results across
two model families and scales from 2B to 8B demonstrate
generalizability without additional training or model-specific
tuning.

On POPE, \methodname improves accuracy for all six model variants
by $0.1$--$1.1$ percentage points and reduces generation steps in
five settings. InternVL3.5-8B improves from $82.8\%$ to $83.9\%$
while reducing steps from $79$ to $76$, and Qwen3-VL-2B-Thinking
improves from $86.8\%$ to $87.2\%$ while reducing steps from
$112$ to $104$. The only exception is InternVL3.5-2B, whose cost
increases marginally from $78$ to $79$ steps while accuracy rises
from $83.6\%$ to $84.5\%$. Overall, the efficiency gains do not
systematically degrade visual grounding and modestly improve
object-hallucination robustness.
\begin{table}[t] 
\captionsetup{skip=3pt}
\centering
\small 
\setlength{\tabcolsep}{18pt} 
\caption{Ablation study on the explicit reasoning maintenance window size ($W$) on the M$^3$CoT benchmark. Our default setting ($W=512$) is highlighted.}
\label{tab:ablation_window}
\begin{tabular}{c cc}
\toprule
\multirow{2}{*}[-0.2em]{Window Size ($W$)} & \multicolumn{2}{c}{M$3$CoT} \\
\cmidrule(lr){2-3} 
& Acc $\uparrow$ & Eff $\downarrow$ \\
\midrule
64   & 52.2 & 209 \\
128  & 58.9 & 412 \\
256  & 63.6 & 609 \\
\rowcolor{blue!10}
512  & \textbf{67.4} & 861 \\ 
1024 & 62.9 & 1766 \\
\bottomrule
\end{tabular}
\end{table}

\subsection{Comparison with LEAD}
\label{sec:additional_baselines}

\begin{table}[h]
  \centering
  \setlength{\tabcolsep}{4pt} 
  \caption{Comparison results with LEAD and our AGS.}
  \label{tab:main_results_single_b}
  \makebox[\columnwidth][c]{
  \resizebox{\columnwidth}{!}{
    \begin{tabular}{lcccccccc}
    \toprule
    \multirow{2}{*}{\textbf{Method}} & 
    \multicolumn{2}{c}{\textbf{V$^\ast$}} & 
    \multicolumn{2}{c}{\textbf{WeMath}} & 
    \multicolumn{2}{c}{\textbf{ScienceQA}} & 
    \multicolumn{2}{c}{\textbf{M$^3$CoT}} \\ 
    \cmidrule(lr){2-3} \cmidrule(lr){4-5} \cmidrule(lr){6-7} \cmidrule(l){8-9} 
     & \textbf{Acc $\uparrow$} & \textbf{Eff $\downarrow$} & \textbf{Acc $\uparrow$} & \textbf{Eff $\downarrow$} & \textbf{Acc $\uparrow$} & \textbf{Eff $\downarrow$} & \textbf{Acc $\uparrow$} & \textbf{Eff $\downarrow$} \\ 
    \midrule
    Explicit CoT & \textbf{82.7} & 462 & 43.1 & 3217 & 81.3 & 1242 & 57.5 & 2255 \\
    LEAD & 80.6 & 413 & \textbf{72.5} & 1706 & 73.6 & 871  & 64.5 & 1511 \\
    \cellcolor{blue!10}AGS (Ours) & \cellcolor{blue!10}81.1 & \cellcolor{blue!10}346 & \cellcolor{blue!10}69.4 & \cellcolor{blue!10}1114 & \cellcolor{blue!10}\textbf{88.2} & \cellcolor{blue!10}643 & \cellcolor{blue!10}\textbf{65.8} & \cellcolor{blue!10}985 \\ 
    \bottomrule
    \end{tabular}
  }}
\end{table}

Table~\ref{tab:main_results_single_b} compares \methodname with
LEAD using the same Qwen3-VL-4B-Thinking checkpoint.
\methodname requires fewer generation steps on all four benchmarks
and achieves higher accuracy on V$^\ast$, ScienceQA, and
M$^3$CoT, while LEAD performs better on WeMath. Specifically, \methodname reduces the cost from $413$ to $346$ steps on
V$^\ast$, from $1706$ to $1114$ on WeMath, from $871$ to $643$ on
ScienceQA, and from $1511$ to $985$ on M$^3$CoT. These gains span fine-grained perception,
scientific reasoning, and general multimodal multi-step reasoning,
demonstrating a more favorable overall accuracy--efficiency
trade-off under the same checkpoint.

\subsection{Ablation Study}
\paragraph{Metric for Reasoning Mode Switching}
We compare the proposed vision-to-text attention ratio with
Explicit CoT and token-level entropy routing. As shown in
Table~\ref{tab:ablation_metric}, entropy routing reduces the average
cost from $1961$ to $864$ steps but provides limited accuracy gains,
consistent with the \textit{uncertainty conflation} issue: entropy
cannot distinguish perceptual ambiguity from logical difficulty and
may therefore select inappropriate reasoning modes. By directly
tracking cross-modal attention flow, our metric better separates
perception-dominant and logic-dominant phases. It improves average
accuracy over entropy routing by $4.0$ percentage points
($70.7\%$ to $74.7\%$) while further reducing the cost to $815$
steps. The gains reach $7.9$ and $5.9$ points on WeMath and
MathVerse, demonstrating a more effective and interpretable routing
signal.

\begin{figure}[t] 
\centering
\includegraphics[width=\columnwidth]{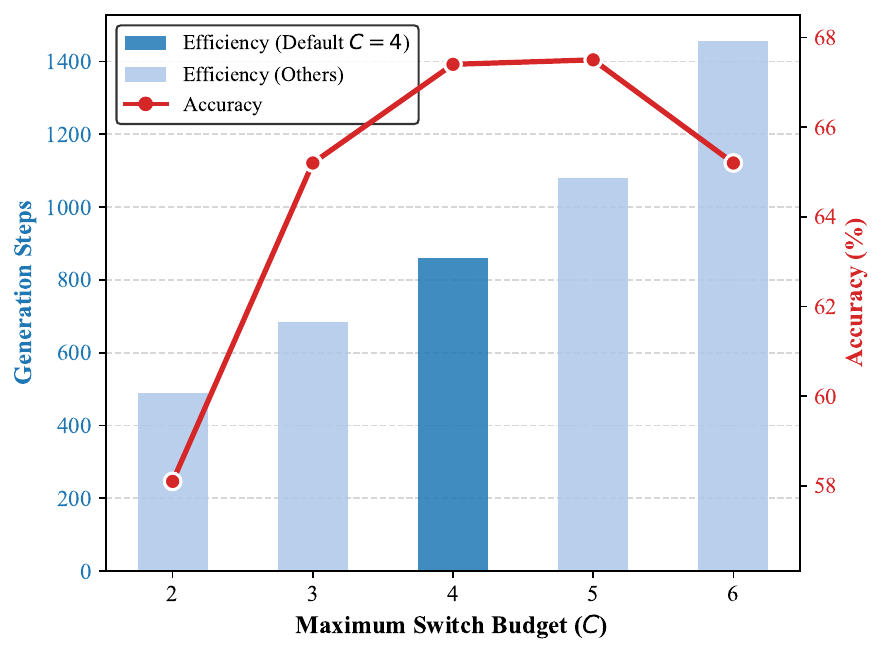} 
\vspace{-0.2cm} 
\caption{Ablation study on the maximum mode transition budget ($C$) using the Qwen3-VL-8B-Thinking model on the M$^3$CoT. The bar chart (left axis) reports inference efficiency in generation steps, while the line plot (right axis) represents reasoning accuracy.}
\label{fig:ablation_budget}
\end{figure}

\begin{figure*}[t]
    \centering
    \includegraphics[width=\textwidth]{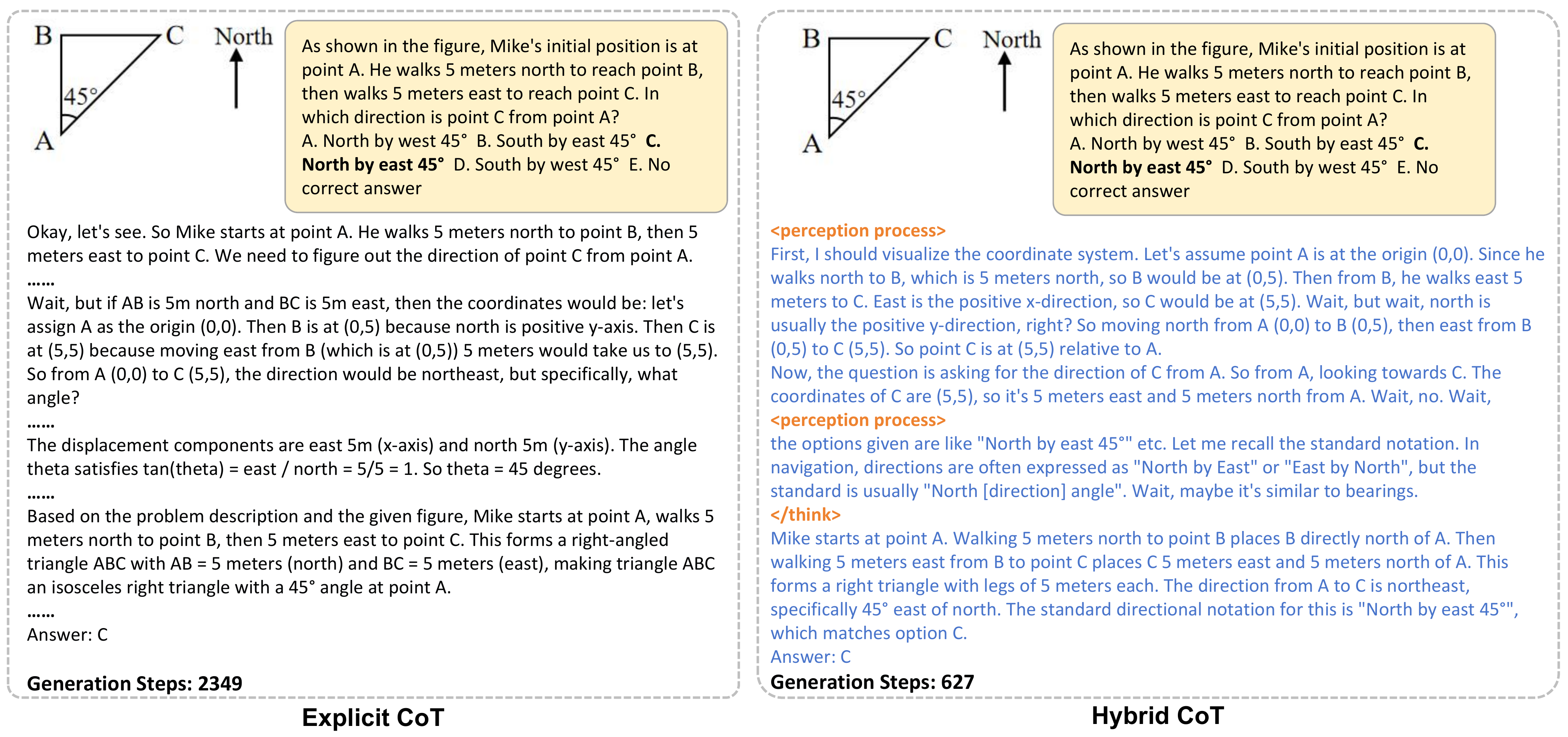}
    \caption{Qualitative comparison between standard Explicit CoT (left) and our proposed Hybrid CoT (right) on a spatial reasoning sample from the WeMath dataset.}
    \label{fig:case_study}
\end{figure*}

\begin{figure}[t]
    \centering
    \includegraphics[width=\linewidth]
    {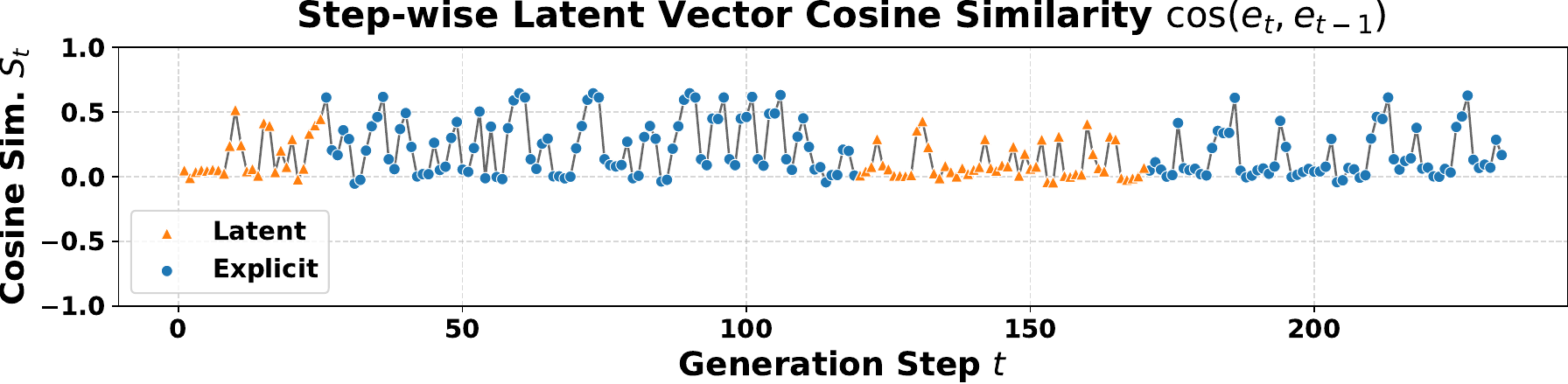}
    \caption{
    Step-wise cosine similarity between consecutive input embeddings,
    $\cos(e_t,e_{t-1})$, for latent and explicit reasoning steps in a
    representative WeMath example using Qwen3-VL-4B-Thinking. Latent similarities remain below $1$ and fluctuate like explicit steps, indicating stable propagation.
    }
    \label{fig:cosine_sim}
\end{figure}

\paragraph{Maintenance Window Size}
We evaluate the sensitivity to the minimum maintenance window
using Qwen3-VL-8B-Thinking on M$^3$CoT, with $W$ ranging from
$64$ to $1024$ generation steps. As shown in
Table~\ref{tab:ablation_window}, performance follows a clear
inverted U-shaped trend. Small windows such as $W=64$ or $128$
provide high efficiency but substantially reduce accuracy, reaching
only $52.2\%$ at $W=64$. Such settings return the model to latent
perception before a coherent explicit deduction is completed,
resulting in fragmented reasoning chains.
Conversely, $W=1024$ keeps the model in explicit generation for
too long, increasing the cost to $1766$ steps while reducing
accuracy to $62.9\%$. This suggests that overly extended textual
reasoning can drift away from continuous visual evidence and become
dominated by verbose linguistic priors. We therefore select
$W=512$, which provides the best balance between coherent logical
anchoring and timely return to high-fidelity visual perception.

\paragraph{Maximum Switch Budget}
We further analyze the effect of the maximum switch budget $C$.
As shown in Figure~\ref{fig:ablation_budget}, accuracy follows an
inverted U-shaped trend, while excessive switching increases
computational cost. With $C=2$, inference requires fewer than
$500$ steps, but accuracy drops to approximately $58\%$, as
insufficient perception--reasoning interactions prevent the model
from resolving complex multi-hop dependencies.
Increasing the budget beyond $C=4$ introduces redundant
transitions without further gains. At $C=6$, accuracy decreases to
around $65\%$, while the generation cost exceeds $1400$ steps,
indicating computational bloat, overthinking, and logical drift.
We therefore set $C=4$, which achieves the highest accuracy of
approximately $67.5\%$ with bounded generation cost.

\subsection{Case Study}
Figure~\ref{fig:case_study} compares Explicit CoT with
\methodname on a WeMath spatial-reasoning example. Explicit CoT
requires $2349$ steps to repeatedly textualize visual coordinates
and displacements, resulting in verbose reasoning and potential
logical loops. Guided by the attention ratio, \methodname routes
visual perception through the continuous latent space
(\texttt{<perception process>}) while retaining explicit generation
for logical deduction. It reaches the correct answer (Option C) in
only $627$ steps, reducing generation cost by approximately $73\%$.
This demonstrates that selective latent routing avoids redundant
visual textualization without sacrificing reasoning accuracy.

\subsection{Stability of Latent Trajectories}
We assess latent propagation using the step-wise cosine similarity
between consecutive input embeddings, $\cos(e_t,e_{t-1})$.
As shown in Figure~\ref{fig:cosine_sim}, latent similarities remain
below $1$ and fluctuate throughout generation, indicating evolving
rather than collapsed representations. Their ranges and variation are
also comparable to explicit steps, suggesting stable representation
dynamics during mode switching.

\section{Conclusions}
\label{sec:conclusion}

In this paper, we identify a key limitation of entropy-based latent
reasoning for MLLMs: perceptual ambiguity and reasoning uncertainty
are entangled. To address this issue, we introduce the interpretable
vision-to-text attention ratio to distinguish perception-dominant
from logic-dominant decoding phases. Based on this metric, our
training-free strategy routes perceptual tokens through the
continuous latent space to preserve visual information, while
retaining explicit generation for structured reasoning. Extensive
experiments across diverse model families and multimodal reasoning
benchmarks show that our method improves reasoning accuracy,
reduces visual hallucinations, and substantially lowers
autoregressive decoding steps and inference latency, without
architectural modification or task-specific supervision. This
provides an efficient alternative to costly model retraining.

\begin{acks}
We want to thank the anonymous reviewers and the meta-reviewer for their valuable comments and suggestions. 
This work is supported in part by the National Science and Technology Major Project 2025ZD1601300, the National Natural Science Foundation of China under Grants 624B2088, 62536003 and 62301189, and by the project of Peng Cheng Laboratory (PCL2025A14).
\end{acks}

\bibliographystyle{ACM-Reference-Format}
\balance
\bibliography{main}

\end{document}